\documentstyle[12pt]{article}

\newcommand{\sect}[1]{\setcounter{equation}{0}\section{#1}
\vspace{-7mm}\indent}
\newcommand{\subsect}[1]{\subsection{{\it #1}}
\vspace{-7mm}\indent}
\newcommand{\subsubsect}[1]{\subsubsection*{#1}
\vspace{-7mm}\indent}

\begin{document}
 
\topmargin -5mm
\oddsidemargin 5mm
 
\begin{titlepage}
\setcounter{page}{0}

\begin{flushright}
OU-HET 252 \\
September 1996
\end{flushright}

\vfill
\begin{center}
\begin{Large} 
 {\bf Topological Field Theory} \\ 
\vspace{-2mm}
  {\bf and}  \\ 
\vspace{2mm}
 {\bf Second-Quantized Five-Branes }
\end{Large}

\vspace{15mm}

{\large 
Kazuyuki Furuuchi 
\footnote{e-mail address 
: furu@phys.wani.osaka-u.ac.jp},
Hiroshi Kunitomo 
\footnote{e-mail address 
: kunitomo@phys.sci.osaka-u.ac.jp} 
\\
and
\\
Toshio Nakatsu
\footnote{e-mail address 
: nakatsu@phys.wani.osaka-u.ac.jp} 
}

\vspace{5mm}
{\em Department of Physics,\\
Graduate School of Science, Osaka University, \\
Toyonaka, Osaka 560, JAPAN}
\end{center}

\vspace{10mm}
\centerline{\bf Abstract}
We construct the six-dimensional topological field 
theory appropriate to describe the ground-state 
configurations of D5-branes. 
A close examination on the degenerations of 
D5-branes gives us the physical observables  
which can be regarded as the Poincar\'e duals 
of the cycles of the moduli space. 
These observables are identified with 
the creation opeartors of the bound states 
of D5-branes and lead to the second quantization of 
five-branes. 
This identification of the bound states with the cycles 
also provides their topological stability and 
suggests that the bound states of five-branes 
have internal structures. 
The partition function of the second-quantized 
five-branes is also discussed.  
 
\vspace{5mm}

\end{titlepage}

\newpage

\sect{Introduction}

              The recent discovery of the 
Dirichlet branes (D-branes) 
\cite{POLCHINSKI} 
gives us a route to 
the quantization of solitonic objects in 
string theory. The quantum fluctuations of these 
solitonic objects are described by 
open strings with one or both of 
their boundaries 
constrained on them, for which they are named 
Dirichlet branes.

          Their low energy effective worldvolume theory 
turns out \cite{PW},\cite{si} 
to be essentially a dimensionally reduced 
supersymmetric gauge theory. 
For the case of the Dirichlet five branes (D5-branes) 
in Type I string theory, 
which are identified 
\cite{si} 
with the dual of the small instanton 
limit of the gauge five branes 
\cite{stro} 
in the $SO(32)$ heterotic 
string, 
the effective worldvolume theory 
is a six-dimensional supersymmetric 
$Sp(1)$ gauge theory. 
It is also argued in 
\cite{si} 
that the $Sp(1)$ gauge symmetry 
can be enhanced to $Sp(k)$ group for $k$ 
coincident D5-branes and that 
the $D$-flat condition gives us  
the ADHM equation 
\cite{ADHMconstruction} of $k$ instantons 
of four-dimensional $SO(32)$ gauge theory.

         In addition to these low energy analysis, 
due to their coupling with open string,    
the D-branes are identified   
with the BPS states which have the Ramond-Ramond charges.  
These BPS states are allowed  
to have the bound states which are marginally stable. 
The existence of these marginally stable bound 
states is argued by several authors 
\cite{bound},\cite{SEN},\cite{VAFA}. 
It is also pointed out \cite{VAFA} the possibility 
that the BPS states can be identified with the 
cohomology elements of instanton moduli spaces.

         In this article  topological aspects 
of the BPS states are studied 
from the worldvolume theoretical viewpoint. 
In particular, 
based on the effective worldvolume gauge theory,  
we examine the geometrical interpretation 
of the bound states.  
For this purpose 
we consider the five branes in the Type IIB theory 
with open strings having 
the $U(1)$ Chan-Paton factors. 
In section 2 we consider the case of 
the $U(n)$ Chan-Paton factors. 
The $D$-flat condition of 
the effective worldvolume theory 
leads to the ADHM equation of 
four-dimensional $U(n)$ gauge theory. 
We construct the six-dimensional 
topological $U(k)$ gauge theory of which physical moduli 
space consists of the solutions of the ADHM equation of 
the $k$ instantons.  
In section 3 we concentrate on the case of 
the $U(1)$ Chan-Paton factors. 
We begin by comparing the physical moduli 
space with the configuration space of D5-branes. 
The degeneration of D5-branes are studied in detail. 
This close examination brings us to introduce 
the physical observable $O_k$ (\ref{Ok}) 
which can be regarded as the Poincar\'e dual of 
the cycle corresponding to $k$ coincident D5-branes.     
These physical observables $O_k$ 
turn out to give a field theoretical realization 
of the correspondences 
investigated by Nakajima \cite{NAKAJIMA},
\cite{GROJNOWSKI} and are 
naturally identified with creation operators of the bound 
states of D5-branes. This identification of the bound states 
with the cycles provides their topological stability 
and suggests that 
the bound states of 5-branes have internal structures.   
The introduction of the creation operators also leads to 
{\em the second quantization} of D5-branes. 
The partition function of second-quantized D5-branes 
gives the generating function of the Poincar\'e polynomials 
of the physical moduli spaces.  
In section 4 we discuss further on the second quantization 
of D5-branes from the four-dimensional supersymmetric 
gauge theoretical point of view.


\sect{Topological Field Theory on D5-Brane Worldvolume}

\subsect{ Effective Worldvolume Theory of D5-brane}

                Let us consider type IIB theory 
with the $U(n)$ open strings\footnote{
This is equivalent to consider the 
type IIB theory under the 
background of $n$ coincident D9-branes\cite{POLCHINSKI}.}.
When there is a D5-brane 
an open string can have either free (Neumann) 
boundary with the (anti-)fundamental 
representation of $U(n)$, 
or Dirichlet boundary located on the  D5-brane.
The Dirichlet boundary has the index of 
the fundamental representation of $U(k)$ 
when there are $k$ coincident D5-branes\cite{bound}. 
The combinations of these boundary conditions  
correspond to three distinct open string sectors: 
Neumann-Neumann (NN), Dirichlet-Dirichlet (DD) 
and Dirichlet-Neumann (DN). 
The quantization of DD and DN strings 
leads to the (first) quantization of 
5-brane\cite{POLCHINSKI},\cite{PW},\cite{si}.

             Suppose that there are 
$k$\ coincident D5-branes at $x^6=\cdots=x^9=0$\ 
for difiniteness. 
Their low energy effective worldvolume theory can 
be described by massless modes 
of the DD and DN strings.
It is a six-dimensional supersymmetric 
$U(k)$\ guage theory with global 
$U(n)$\ symmetry\cite{PW},\cite{si}.

                       There are two kinds 
of massless bosonic modes 
$A_{\mu}$\ and $X^i$\ in the DD sector. 
$A_{\mu}\ (\mu=0,\cdots,5)$\ give 
a $U(k)$\ gauge field. 
$X^i\ (i=6,\cdots,9)$\ are scalar fields which 
belong to the adjoint representations of $U(k)$. 
Their $U(k)$\ gauge transformations are 
\begin{eqnarray} 
A_{\mu}(x)&\rightarrow& 
g(x)A_{\mu}(x)g^{-1}(x)-i(\partial_{\mu}g(x))g^{-1}(x),\\
X^i(x)&\rightarrow& g(x)X^i(x) g^{-1}(x), 
\qquad\qquad g(x)\in U(k). 
\end{eqnarray}
Note that $A_{\mu}$ and $X^i$ 
are represented 
by $k\times k$ hermitian matrices. 
These fields are inert under the global $U(n)$ 
since the DD string has no Neumann boundary. 
The vacuum expectation values of $X^i$ 
($6 \leq i \leq 9$) become 
the collective coordinates 
of D5-branes. 
The $SO(4)\simeq SU(2)\times SU(2)_R$ 
which orginates from 
the rotations 
in the four-dimensions 
$(x^6,\cdots,x^9)$ 
is a global symmetry group 
of the worldvolume theory. 
$SU(2)_R$ is identified 
with the $SU(2)$ R-symmetry. 
In order to make it clear, 
it is convenient to rewrite $X^i$ as 
\begin{equation}
X_{A \dot{A}}=X^i\sigma^i_{A\dot{A}},\qquad 
\sigma^i_{A\dot{A}}=(i\tau^1,i\tau^2,i\tau^3,{\bf 1}_2),\\
\end{equation}
where $\tau^{1,2,3}$\ are the Pauli matricies  
and ${\bf 1}_2$ is a $2\times 2$ identity matrix. 
$\dot{A}(=\dot 1,\dot 2)$ is the $SU(2)_R$ index and 
$A(=1,2)$ is the $SU(2)$ index. 
In the DN sector, 
there is a $SU(2)_R$ doublet complex scalar $H_{\dot{A}}$. 
Since the DN string has both Neumann and Dirichlet 
boundaries, each $H_{\dot{A}}$ transforms as 
$({\bf k}, \bar{\bf n})$ representation 
under the action of $U(k)\times U(n)$.
\begin{eqnarray} 
H_{\dot{A}}(x)&\rightarrow& g(x)H_{\dot{A}}(x) h^{-1},
\qquad\qquad 
(g(x),h)\in U(k)\times U(n). 
\end{eqnarray} 
Notice that $H_{\dot{A}}$ is represented 
by a $k\times n$\ complex matrix.

               The bosonic part of 
the effective action will be given by 
\begin{eqnarray}
S_{boson}
&=&\int \! d^6x~ 
{\rm Tr} \left[
  {1\over 4}F^{\mu\nu}F_{\mu\nu}
       +{1\over 2}D^{\mu}X^{A\dot{A}}
                       D_{\mu}X_{A\dot{A}}
       +D^{\mu}H_{\dot{A}}D_{\mu}
                          \bar{H}^{\dot{A}} \right. 
\nonumber\\ 
&&\qquad\qquad
 \left.    
   -{1\over 2}D^{\dot{A}\dot{B}}D_{\dot{A}\dot{B}}
            +D^{\dot{A}\dot{B}}
(X_{A \dot{A}} X^A_{\dot{B}}+H_{\dot{A}}\bar{H}_{\dot{B}})
\right],
\label{effective}
\end{eqnarray}
where $D_{\mu}$ is 
the $U(k)$ gauge covariant 
derivative and 
\begin{eqnarray}
\bar{H}_{\dot{A}}&=&
\epsilon_{\dot{A}\dot{B}}(H_{\dot{B}})^{\dag}.
\end{eqnarray}
The degenerate vacua (moduli space) 
are determined by the D-flat conditon
\begin{equation}
X_{A(\dot{A}} X^A_{\dot{B})}+H_{(\dot{A}}\bar{H}_{\dot{B})}=0 ,
\label{Dflat}
\end{equation}
which coincides with the ADHM equation 
\cite{ADHMconstruction}
of $U(n)$ instantons in ${\bf R}^4$.

                   So far we have implicitly assumed that 
the space-time is flat ${\bf R}^{10}$, 
but we may consider more general cases.  
For the case that 
the four-dimensional space $(x^6,\cdots,x^9)$ 
is an ALE space $K$, 
ADHM equation (\ref{Dflat}) 
will be modified \cite{KN} to 
\begin{equation}
X_{A(\dot{A}} X^A_{\dot{B})}+H_{(\dot{A}}\bar{H}_{\dot{B})}
=\zeta_{\dot{A}\dot{B}}{\bf 1}_k,
\label{ADHM}
\end{equation}
where ${\bf 1}_k$ is a $k\times k$ unit matrix and
$\zeta_{\dot{A}\dot{B}} (=\zeta_{\dot{B}\dot{A}}) $ are the constants 
related to the three Kh\"aler forms 
$\omega_{\dot{A}\dot{B}}$ of $K$ : 
$\zeta_{\dot{A}\dot{B}}
=\int_{C_2}\omega_{\dot{A}\dot{B}},~ C_2\in H_2(K)$. 
This modification of the ADHM equation 
can be also understood 
from the D-brane viewpoint 
as the result of 
the Fayet-Iliopoulos D-terms 
induced in a non-trivial 
gravitational background \cite{quiver} : 
\begin{equation}
S_{FI}=-\zeta_{\dot{A}\dot{B}}\int \! d^6x
~{\rm Tr}D^{\dot{A}\dot{B}}.
\label{FI}
\end{equation}
Notice that, 
by using the $SU(2)_R$ rotations, 
we may set these three constants 
$\zeta_{\dot{A}\dot{B}}$ equal to zero 
except the only one component 
which we call $\eta (>0)$. 
This can be done when we fix a complex structure
and the constant $\eta$ is related 
to the corresponding K\"ahler form 
$\omega^R$ by 
$\eta=\int_{C_2}\omega^R$.

        As the first step to taking account of 
the gravitational effect (due to string) 
it may be enough to add Fayet-Iliopoulos 
D-terms (\ref{FI}) to the effective action. 
(The corresponding $\eta$ is a positive constant.)
Therefore the moduli space of the effective worldvolume 
$U(k)$ gauge theory will be given as  
\vspace{3mm}
\begin{equation}
{\cal M}^k_{U(n)}
=
\left. \left\{ 
(X_{A \dot{A}}, H_{\dot{A}})~ 
\Biggl| ~ \Biggr. 
{
 X_{A(\dot{A}} X^A_{\dot{B})}
+H_{(\dot{A}}\bar{H}_{\dot{B})}
=\zeta_{\dot{A}\dot{B}}{\bf 1}_k,
\atop
\partial_{\mu}X_{A \dot{A}}=\partial_{\mu}H_{\dot{A}}=0\hfill
}
\right\} \right/ U(k),
\label{moduli}
\end{equation}

\vspace{3mm}
\noindent
where the subscript  $U(n)$ 
denotes the Chan-Paton factors of the 
open strings. 
The quotient by $U(k)$ in the R.H.S. 
of (\ref{moduli}) 
must be taken due to the global part 
of the $U(k)$ gauge symmetry. 
As regards the dimensionality of the moduli 
space, by simply counting up the degrees of freedom, 
it turns out to be 
\begin{eqnarray}
\mbox{dim}_{{\bf R}}
{\cal M}^k_{U(n)}=4kn. 
\label{dim of M}
\end{eqnarray} 
Notice that the moduli space 
${\cal M}^k_{U(n)}$ 
also describes 
the grand-state configurations 
of D5-branes in Type IIB theory 
with the $U(n)$ open strings.

\subsect{Topological Field Theory}

                     Let us construct a six-dimensional 
topological gauge theory appropriate 
to describe the cohomology 
theory on moduli space 
${\cal M}^k_{U(n)}$ (\ref{moduli}). 
To obtain a topological field theory \cite{tft} 
it may be convenient to use 
the standard BRST formulation 
\cite{BRSTformalism}.  
To apply this formalism we shall begin by considering 
the following constraints on the field variables :
\begin{eqnarray}
F_{\mu\nu}&\equiv&
\partial_{\mu}A_{\nu}
-\partial_{\nu}A_{\mu}-i[A_{\mu},A_{\nu}] 
\nonumber \\ 
&=&0 , 
\label{constraintb}\\
D_{\mu}X_{A \dot{A}} &\equiv& 
\partial_{\mu}X_{A \dot{A}}-i[A_{\mu},X_{A \dot{A}}] 
\nonumber \\ 
&=&0  ,
\label{constraintc} \\
D_{\mu}H_{\dot{A}} 
&\equiv& \partial_{\mu}H_{\dot{A}}-iA_{\mu}H_{\dot{A}} 
\nonumber \\
&=&0, 
\label{constraintd} \\
X_{A(\dot{A}} X^A_{\dot{B})}
+H_{(\dot{A}}\bar{H}_{\dot{B})}&=&
\zeta_{\dot{A}\dot{B}}{\bf 1}_k . 
\label{constrainta}
\end{eqnarray}
Since the worldvolume is ${\bf R}^6$ 
one can solve the first constraint  
$F_{\mu \nu}=0$ by simply setting $A_{\mu}$ a pure gauge. 
Constarints (\ref{constraintc}) and (\ref{constraintd}) 
now reduce to  
\begin{equation}
\partial_{\mu}X_{A \dot{A}}=\partial_{\mu}H_{\dot{A}}=0,
\end{equation}
which means that $X_{A \dot{A}}$ and 
$H_{A \dot{A}}$ are constant. 
Therefore the last constraint becomes equivalent to  
ADHM equation (\ref{ADHM}).

                 As the second step we define 
the (pre-)BRST transformation 
$\tilde{\delta}$ by introducing the 
corresponding (fermionic) ghost fields 
$\psi_{\mu},\psi_{A \dot{A}}$ and $\psi_{\dot{A}}$ :  
\begin{eqnarray}
\tilde\delta A_{\mu}&=&\psi_{\mu},\qquad
\tilde\delta \psi_{\mu}=0,\nonumber \\
\tilde\delta X_{A \dot{A}}&=&\psi_{A\dot{A}},\qquad
\tilde\delta \psi_{A\dot{A}}=0,\nonumber \\
\tilde\delta H_{\dot{A}}&=&\psi_{\dot{A}},\qquad
\tilde\delta \psi_{\dot{A}}=0 .
\end{eqnarray}
To construct the action of topological 
field theory one may need 
(bosonic) auxiliary fields and 
(fermionic) anti-ghost fields. 
Let us introduce 
$D_{\dot{A}\dot{B}}, \chi_{\mu\nu},\chi_{\mu A\dot{A}}$  
and $\chi_{\mu \dot{A}}$   
as the auxiliary fields. 
The anti-ghost fields are denoted by 
$\eta_{\dot{A}\dot{B}}, \eta_{\mu\nu},
\eta_{\mu A\dot{A}}$ 
and $\eta_{\mu \dot{A}}$.
The (pre-)BRST transforms of these additional fields 
are given by  
\begin{eqnarray}
\tilde\delta \eta_{\dot{A}\dot{B}}&=&iD_{\dot{A}\dot{B}},
\qquad
\tilde\delta D_{\dot{A}\dot{B}}=0,
\nonumber \\
\tilde\delta \eta_{\mu\nu}&=&i\chi_{\mu\nu},
\qquad
\tilde\delta \chi_{\mu\nu}=0,
\nonumber \\
\tilde\delta \eta_{\mu A\dot{A}}&=&i\chi_{\mu A\dot{A}},
\qquad
\tilde\delta \chi_{\mu A\dot{A}}=0,
\nonumber \\
\tilde\delta \eta_{\mu \dot{A}}&=&i\chi_{\mu \dot{A}},
\qquad
\tilde\delta \chi_{\mu \dot{A}}=0 .
\label{preBRST}
\end{eqnarray}
Notice that these (anti-)ghost and auxiliary fields 
are $k \times k$ or $k \times n$ matrix-valued 
with the definite values of ghost number. 
For instance the fermionic ghost 
$\psi_{A \dot{A}}$ is 
$k \times k$ hermitian matrix-valued 
and possesses the ghost number equal to 
one. 
Matrix sizes and ghost numbers 
of other fields are summarized 
in Table 1. 


                  Now the $\tilde{\delta}$-invariant 
action will follow if we apply the standard BRST 
formalism with taking constraints 
(\ref{constraintb})-(\ref{constrainta}) 
as the gauge-fixing functionals : 
\begin{eqnarray}
\widetilde S 
&=&
-i\int \!d^6x~ 
      \tilde\delta {\rm Tr} 
\left[{1\over 2}\eta^{\mu\nu}F_{\mu\nu}
       +\eta^{\mu A\dot{A}}D_{\mu}X_{A \dot{A}}
          +\eta^{\mu}_{\dot{A}}D_{\mu}\bar{H}^{\dot{A}}
             +(D_{\mu}H_{\dot{A}})\bar\eta^{\mu\dot{A}} \right. 
\nonumber \\
        &&\qquad
\left.+\eta^{\dot{A}\dot{B}}(X_{A \dot{A}} X^A_{\dot{B}}
            +H_{\dot{A}}\bar{H}_{\dot{B}}
                  -\zeta_{\dot{A}\dot{B}}{\bf 1}_k) \right],  
\end{eqnarray} 
which is evaluated into the form : 
\begin{eqnarray} 
\widetilde S 
  &=&\int \!d^6x~ {\rm Tr}
  \left[{1\over 2}\chi^{\mu\nu}F_{\mu\nu}
             +\chi^{\mu A\dot{A}}D_{\mu}X_{A \dot{A}}
     +\chi^{\mu}_{\dot{A}}D_{\mu}\bar{H}^{\dot{A}}
     +(D_{\mu}H_{\dot{A}})\bar\chi^{\mu\dot{A}} \right. 
\nonumber \\
  &&\qquad
       +D^{\dot{A}\dot{B}}(X_{A \dot{A}} X^A_{\dot{B}}
              +H_{\dot{A}}\bar{H}_{\dot{B}}
             -\zeta_{\dot{A}\dot{B}}{\bf 1}_k)
\nonumber \\
  &&\qquad
      +i\eta^{\mu\nu}D_{\mu}\psi_{\nu}
                +i\eta^{\mu A\dot{A}}
                    D_{\mu}\psi_{A\dot{A}}
     +i\eta^{\mu}_{\dot{A}}D_{\mu}\bar\psi^{\dot{A}}
       -i(D_{\mu}\psi_{\dot{A}})\bar\eta^{\mu\dot{A}}
\nonumber \\
  &&\qquad
      \left. +i\eta^{\dot{A}\dot{B}}(\psi_{A\dot{A}}X^A_{\dot{B}}
                        +X_{A \dot{A}}\psi^A_{\dot{B}}
       +\psi_{\dot{A}}\bar{H}_{\dot{B}}
                        +H_{\dot{A}}\bar\psi_{\dot{B}}) \right].
\label{wideS}
\end{eqnarray}
Note that the equations of motion for $A_{\mu}, X_{A\dot{A}}$ 
and $H_{\dot{A}}$ 
fields are exactly equations 
(\ref{constraintb})-(\ref{constrainta}) 
and that those for the ghost fields 
have the forms :  
\begin{eqnarray}
&&D_{[\mu}\psi_{\nu]}=0,  
\label{ghostb} \\
&&D_{\mu}\psi_{A\dot{A}}=0,
\label{ghostc} \\
&&D_{\mu}\psi_{\dot{A}}=0,
\label{ghostd} \\
&&\psi_{A(\dot{A}}X^A_{\dot{B})}+X_{A(\dot{A}}\psi^A_{\dot{B})}
+\psi_{(\dot{A}}\bar{H}_{\dot{B})}
+H_{(\dot{A}}\bar\psi_{\dot{B})}=0.
\label{ghosta}
\end{eqnarray}
Since the worldvolume is ${\bf R}^6$ one can also solve 
equations (\ref{ghostb})-(\ref{ghostd}) 
by simply gauging away the nonzero modes of the ghosts 
(due to the triviality of $A_{\mu}$). Last equation 
(\ref{ghosta}) can be regarded as the equations of motion for 
the zero modes. 
Since it describes   
the (infinitesimal) deformation of the ADHM equation, 
the ghost zero modes can be thought of as the (co-)tangent 
vectors
\footnote{Strictly speaking the (co-)tangent vectors 
are given by the equivalent classes of these zero modes 
under the action of $U(k)$.} of moduli space 
${\cal M}^k_{U(n)}$ (\ref{moduli}).

                The above construction of topological field 
theory, however, is not complete yet. 
Some modifications are needed 
due to the existence of two residual local symmetries 
in  action (\ref{wideS}). 
One is the \lq\lq super" gauge symmetry \cite{tft}. 
The transforms of $\psi_{\mu}$ and $\chi_{\mu \nu}$ under 
this fermionic symmetry have the forms : 
\begin{eqnarray}
\delta \psi_{\mu}=D_{\mu}\theta,~~~~
\delta \chi_{\mu \nu}= 
\mbox{[$\eta_{\mu \nu},~\theta$]}, 
\label{addsuper}
\end{eqnarray} 
where $\theta$ is the fermionic gauge parameter. 
The another residual symmetry is rather peculiar, which 
transforms the auxiliary and anti-ghost fields by 
anti-symmetric tensorial parameters 
$\Lambda_{\mu \nu \cdots}$, for instance, 
\begin{eqnarray}
\delta \chi^{\mu \nu}
=\frac{1}{3!}\epsilon^{\mu \nu \rho_1 \rho_2 
  \rho_3 \rho_4 }D_{\rho_1}
  \Lambda_{\rho_2 \rho_3 \rho_4} , 
\label{tensorsymm}
\end{eqnarray}
where $\Lambda_{\mu \nu \rho}$ is 
the rank three anti-symmetric tensorial parameter. 
This peculiar symmetry originates in 
the properties of equations 
(\ref{constraintb})-(\ref{constrainta}). 
For example  symmetry (\ref{tensorsymm}) 
is due to the 
Bianchi identity, 
$D_{\mbox{[}\mu}F_{\nu \rho \mbox{]}}=0$.

                The prescription for  
\lq\lq super" gauge symmetry (\ref{addsuper}) 
is as follows. Let us introduce 
the (bosonic) ghost $\phi$. 
The corresponding (bosonic) anti-ghost 
and (fermionic) auxiliary fields are denoted by  
$\lambda$ and $\eta$ respectively. 
We will fix fermionic symmetry 
(\ref{addsuper}) by imposing the condition :  
\begin{eqnarray}
D_{\mu}\psi^{\mu}
&\equiv& 
\partial_{\mu}\psi^{\mu} 
-i \mbox{[}
A_{\mu},~\psi^{\mu} \mbox{]} 
\nonumber \\
&=& 0.
\end{eqnarray}
(pre-)BRST transforms (\ref{preBRST}) must be 
modified  so that they are consistent with the above   
prescription. 
This requirement leads to the 
following definition 
of the BRST transformation $\delta_{\bf b}$ : 
\begin{eqnarray}
\delta_{\bf b} A_{\mu}&=&\psi_{\mu}, 
\qquad
\delta_{\bf b} \psi_{\mu}=D_{\mu}\phi, 
\nonumber \\
\delta_{\bf b} X_{A \dot{B}}&=&\psi_{A \dot{B}}, 
\qquad
\delta_{\bf b} \psi_{A\dot{A}}=i[\phi,X_{A \dot{A}}], 
\nonumber \\
\delta_{\bf b} H_{\dot{A}}&=&\psi_{\dot{A}}, 
\qquad
\delta_{\bf b} \psi_{\dot{A}}=i[\phi,H_{\dot{A}}], 
\nonumber \\
\delta_{\bf b} \phi&=&0, 
\nonumber \\
\delta_{\bf b} \eta_{\mu\nu}&=&i\chi_{\mu\nu}, 
\qquad
\delta_{\bf b} \chi_{\mu\nu}=[\phi,\eta_{\mu\nu}], 
\nonumber\\
\delta_{\bf b} \eta_{\mu A\dot{A}}&=&i\chi_{\mu A\dot{A}}, 
\qquad
\delta_{\bf b} \chi_{\mu A\dot{A}}=[\phi,\eta_{\mu A\dot{A}}], 
\nonumber\\ 
\delta_{\bf b} \eta_{\mu\dot{A}}&=&i\chi_{\mu\dot{A}},
\qquad
\delta_{\bf b} \chi_{\mu\dot{A}}=[\phi,\eta_{\mu\dot{A}}],
\nonumber\\ 
\delta_{\bf b} \eta_{\dot{A}\dot{B}}&=&iD_{\dot{A}\dot{B}},
\qquad
\delta_{\bf b} D_{\dot{A}\dot{B}}=[\phi,\eta_{\dot{A}\dot{B}}],
\nonumber\\
\delta_{\bf b} \lambda&=&\eta,\qquad
\delta_{\bf b} \eta=i[\phi,\lambda] .
\label{BRST}
\end{eqnarray}
Note that the transformation 
$\delta_{\bf b}$ is nilpotent up to
the $U(k)$\ gauge transformation 
parametrized by the bosonic ghost $\phi$ 
: $~\delta_{\bf b}^2=\delta_{U(k)}(\phi)$.

               As regards the residual symmetry of 
the anti-ghost and auxiliary fields it turns out that 
one can fix it by simply introducing non-zero gauge parameters 
in the action. Taking into account these two modifications 
we finally obtain the following form of the action : 
\begin{eqnarray}
S&=&-i\int \! d^6x 
          ~\delta_{\bf b} {\rm Tr} \left[
       {1\over 2}\eta^{\mu\nu}F_{\mu\nu}
                  +\lambda D^{\mu}\psi_{\mu}
                      +\eta^{\mu A\dot{A}}D_{\mu}X_{A\dot{A}}
           +\eta^{\mu}_{\dot{A}}D_{\mu}\bar{H}^{\dot{A}}
                +(D_{\mu}H_{\dot{A}})\bar\eta^{\mu\dot{A}} \right. 
\nonumber\\
&&\qquad
      +\eta^{\dot{A}\dot{B}}
           (X_{A \dot{A}} X^A_{\dot{B}}
               +H_{\dot{A}}\bar{H}_{\dot{B}}
                     -\zeta_{\dot{A}\dot{B}}I_k)
\nonumber\\
&&\qquad
       \left. -{\alpha_1\over 4}
             \eta^{\mu\nu}\chi_{\mu\nu}
      -{\alpha_2\over 2}
              \eta^{\mu A\dot{A}}\chi_{\mu A\dot{A}}
      -{\alpha_3\over 2}
             \big(\eta_{\mu\dot{A}}\bar\chi^{\mu\dot{A}}
      +\chi_{\mu\dot{A}}\bar\eta^{\mu\dot{A}}\big)
          -{\alpha_4\over 2}
             \eta^{\dot{A}\dot{B}}D_{\dot{A}\dot{B}}
              \right]
\nonumber \\ 
&&~~~
\nonumber \\
&=&
\int \! d^6x~{\rm Tr}\left[ 
{1\over 2}\chi^{\mu\nu}F_{\mu\nu}
-{\alpha_1\over 4}\chi^{\mu\nu}\chi_{\mu\nu}
+\chi^{\mu A\dot{A}}D_{\mu}X_{A \dot{A}}
-{\alpha_2\over 2}\chi^{\mu A\dot{A}}\chi_{\mu A\dot{A}} \right. 
\nonumber\\
&&\qquad
+\chi^{\mu}_{\dot{A}}D_{\mu}\bar{H}^{\dot{A}}
+(D_{\mu}H_{\dot{A}})\bar\chi^{\mu\dot{A}}
-\alpha_3\chi_{\mu\dot{A}}\bar\chi^{\mu\dot{A}}\nonumber\\
&&\qquad
+D^{\dot{A}\dot{B}}(X_{A \dot{A}} X^A_{\dot{B}}
+H_{\dot{A}}\bar{H}_{\dot{B}}-\zeta_{\dot{A}\dot{B}}I_k)
-{\alpha_4\over 2}D^{\dot{A}\dot{B}}D_{\dot{A}\dot{B}} 
\nonumber\\ 
&&\qquad 
+i\eta^{\mu\nu}D_{\mu}\psi_{\nu}
+i\eta^{\mu A\dot{A}}D_{\mu}\psi_{A\dot{A}}
+i\eta^{\mu}_{\dot{A}}D_{\mu}\bar\psi^{\dot{A}}
-i(D_{\mu}\psi_{\dot{A}})\bar\eta^{\mu\dot{A}}\nonumber\\
&&\qquad
+i\eta^{\dot{A}\dot{B}}
(\psi_{A\dot{A}}X^A_{\dot{B}}+X_{A \dot{A}}\psi^A_{\dot{B}}
+\psi_{\dot{A}}\bar{H}_{\dot{B}}+H_{\dot{A}}\bar\psi_{\dot{B}})
\nonumber\\
&&\qquad
+\eta^{\mu A\dot{A}}[\psi_{\mu},X_{A \dot{A}}]
+\eta^{\mu}_{\dot{A}}[\psi_{\mu},\bar{H}^{\dot{A}}]
-[\psi_{\mu},H_{\dot{A}}]\bar\eta^{\mu\dot{A}}\nonumber\\
&&\qquad
-i{\alpha_1\over 4}\eta^{\mu\nu}[\phi,\eta_{\mu\nu}]
-i{\alpha_2\over 2}\eta^{\mu A\dot{A}}[\phi,\eta_{\mu A\dot{A}}]
-i\alpha_3\eta_{\mu\dot{A}}[\phi,\bar\eta^{\mu\dot{A}}]
\nonumber \\ 
&&\qquad
\left. 
-i{\alpha_4\over 2}\eta^{\dot{A}\dot{B}}
[\phi,\eta_{\dot{A}\dot{B}}]  
-i\eta D^{\mu}\psi_{\mu}+iD^{\mu}\lambda D_{\mu}\phi
-\lambda [\psi^{\mu},\psi_{\mu}] 
\right]
\label{lag}
\end{eqnarray}
where $\alpha_1,\cdots,\alpha_4$ are 
the gauge parameters which 
fix the residual symmetry of the anti-ghost 
and auxiliary fields.

          Let us describe the relation between 
the action of the topological theory and that 
of the effective D5-brane worldvolume theory. 
Taking the Feynmann gauge $\alpha_1=\cdots=\alpha_4=1$\ and 
then integrating out the auxiliary fields $\chi$ in (\ref{lag}), 
we can find out that 
the action has the form :  
\begin{eqnarray}
S&=&
\int \! d^6x~{\rm Tr}
\left[ 
{1\over 4}F^{\mu\nu}F_{\mu\nu}
+{1\over 2}D^{\mu}X^{A\dot{A}}D_{\mu}X_{A\dot{A}}
+D^{\mu}H_{\dot{A}}D_{\mu}\bar{H}^{\dot{A}} \right. 
\nonumber\\
&&\qquad\qquad
\left. -{1\over 2}D^{\dot{A}\dot{B}}D_{\dot{A}\dot{B}}
+D^{\dot{A}\dot{B}}(X_{A \dot{A}} X^A_{\dot{B}}+H_{\dot{A}}
\bar{H}_{\dot{B}}-\zeta_{\dot{A}\dot{B}}I_k)
+\cdots 
\right], 
\end{eqnarray}
where $\cdots$\ denotes the terms including fermions.
This is nothing but the bosonic part of action 
(\ref{effective})
of the effective D5-brane worldvolume theory 
with  Fayet-Iliopoulos D-terms (\ref{FI}). 
Therefore our topological field theory can be regarded 
as the one describing the ground-state configurations 
of D5-branes.

             In the next section, 
concentrating on the simple case that 
the open strings have the $U(1)$ Chan-Paton factors, 
we will study the moduli space 
${\cal M}^k$ $(\equiv {\cal M}^k_{U(1)})$ 
from the D5-brane viewpoint 
and then, returning to the topological 
field theory, 
we will construct physical observables. 
These observables will be interpreted  
in terms of D5-branes.

\sect{Second Quantization of D5-Branes}

\subsect{Degenerations of D5-Branes}

          In this section we study D5-branes in 
Type IIB theory with the $U(1)$ open strings. 
We will begin by describing 
the relation between the moduli space 
${\cal M}^k$ $(\equiv {\cal M}^k_{U(1)})$ 
and the configuration space of D5-branes.
The configuration space
of $k$ D5-branes 
is the $k$-th symmetric product of ${\bf R}^4$, 
$S^k({\bf R}^4)$
\footnote{We assume that quantum statistics  
of D5-branes are bosonic.}. 
In fact, when the j-th D5-brane is located on 
$P_j \in {\bf R}^4$($1 \leq j \leq k$),
the corresponding configuration is represented as   
$P_1+ \cdots +P_k \in S^k({\bf R}^4)$.
Let us consider the $U(k)$-invariant 
combinations of 
$(X_{A \dot{A}},H_{\dot{A}})$. 
The characteristic polynomial of 
$X_{A\dot{A}}$ suffices :  
\begin{eqnarray} 
\mbox{det}(\lambda -X_{A\dot{A}})=
\prod_{j=1}^k(\lambda -x_{j,A\dot{A}}).
\end{eqnarray}
Each eigenvalue $x_{j,A\dot{A}}$ can be thought of 
as the coordinates of 
$P_j \in {\bf R}^4$ on which the j-th D5-brane 
is located. 
Therefore this set of eigenvalues 
of $X_{A\dot{A}}$, 
counting their multiplicity if they degenerate,  
defines the projection $\pi$ 
from the moduli space  
${\cal M}^k$ to the configuration space of 
D5-branes. 
\begin{eqnarray}
\pi~:~~
{\cal M}^k \rightarrow S^k({\bf R}^4)~~~~
(X_{A\dot{A}}, H_{\dot{A}})\mapsto 
P_1+\cdots +P_k 
\label{pi}
\end{eqnarray}     
When the positions of $k$ D5-branes are generic, 
$x_{j,A\dot{A}} ~(1 \leq j \leq k)$ 
can be regarded as 
the coordinates of moduli space ${\cal M}^k$ 
(\ref{moduli}).

        We will consider the degeneration of 
D5-branes in  moduli space ${\cal M}^k$ 
(\ref{moduli}).
In particular we will study the case 
when all the D5-branes are 
overlapping at a point. 
For this purpose it is convenient to use the 
center-of-mass and relative coordinates of $k$ D5-branes. 
The center of $k$ D5-branes is described by 
$x_{A \dot{A}}^{(0)}$ :  
\begin{eqnarray}
x_{A\dot{A}}^{(0)}
=\frac{1}{k}\mbox{Tr} X_{A\dot{A}}, 
\end{eqnarray} 
and the relative motions are measured by 
$x_{A \dot{A}}^{(j)}$ : 
\begin{eqnarray} 
x_{A\dot{A}}^{(j)}
=x_{j,A\dot{A}}-x_{A\dot{A}}^{(0)} 
~~~~~~~(1 \leq j \leq k-1). 
\end{eqnarray}  
We  denote  the relative distances 
of $k$ D5-branes by 
$\Delta^{(j)}$ : 
\begin{eqnarray}
\Delta^{(j)~ 2}
= \mbox{det}~x_{A\dot{A}}^{(j)}~~~~~~~~
(1 \leq j \leq k-1). 
\end{eqnarray}

           It also turns out useful 
to introduce the following complex structure  
in  the moduli space ${\cal M}^k$ :  
\begin{eqnarray}
z^{(j)}_{1}=x^{(j)}_{2 \dot{1}},~~~
z^{(j)}_{2}=x^{(j)}_{1 \dot{1}}~~~
(0 \leq j\leq k-1),
\label{cpx}
\end{eqnarray} 
which is equivalent to use  the following 
complex matrices 
at the level of field variables :  
\begin{eqnarray}
N=X_{2 \dot{1}},~~~~~M=X_{1 \dot{1}}~~~.
\label{cpxmatrix}
\end{eqnarray}
In terms of these variables 
the ADHM equation has the form : 
\begin{eqnarray}
\mbox{$[$}~N,~M ~\mbox{$]$}
+H_{\dot{1}}H_{\dot{2}}^{\dagger}
&=&0, 
\nonumber \\
\mbox{$[$}~N,~N^{\dagger}~ \mbox{$]$}
+ \mbox{$[$}~M,~M^{\dagger}~ \mbox{$]$}
+H_{\dot{1}}H_{\dot{1}}^{\dagger}
-H_{\dot{2}}H_{\dot{2}}^{\dagger}
&=&\eta {\bf 1}_k~~~. 
\label{cpxADHM}
\end{eqnarray}

\subsubsect{$(i)~~k=2~~case$}

        We firstly consider the degeneration 
of two D5-branes. 
Their center is located at 
$z^{(0)}=(z^{(0)}_1,z^{(0)}_2)$ 
and the relative coordinate is given by 
$z^{(1)}=(z^{(1)}_1,z^{(1)}_2)$. 
The configuration of these two D5-branes 
being separate from each other gives 
the following solution of ADHM equation 
(\ref{cpxADHM}) 
up to the $U(2)$ action : 
\begin{eqnarray}
N&=& 
\left( \begin{array}{cc}
   z_1^{(0)}+z_1^{(1)}/2 & 
      \frac{\eta}{u}\sqrt{\eta (2-\frac{\eta}{u^2})} 
          \frac{z_1^{(1)}}{\Delta^{(1)}} 
\\ 
0 & z_1^{(0)}-z_1^{(1)}/2 
\end{array} 
\right) ,
\nonumber \\ 
M&=& 
\left( \begin{array}{cc}
   z_2^{(0)}+z_2^{(1)}/2 & 
      \frac{\eta}{u}\sqrt{\eta (2-\frac{\eta}{u^2})} 
          \frac{z_2^{(1)}}{\Delta^{(1)}} 
\\ 
0 & z_2^{(0)}-z_2^{(1)}/2 
\end{array} 
\right) ,
\nonumber \\
H_{\dot{1}}&=& 
\left( \begin{array}{c} 
   \eta /u \\ 
\sqrt{\eta(2-\frac{\eta}{u^2})} 
\end{array} 
\right),~~~
H_{\dot{2}}=
\left( \begin{array}{c} 
  0 \\ 0 
\end{array} 
\right),  
\label{k=2solution1}
\end{eqnarray}
where 
\begin{eqnarray}
u^2=
\frac{\eta}{2}\left\{ 
1+\frac{2\eta}{\Delta^{(1)~2}} 
+\sqrt{1+\left( \frac{2\eta}{\Delta^{(1)~2}} 
\right)} \right\}.
\label{k=2u}
\end{eqnarray}

        The collision of two D5-branes may 
be described by 
studying the behavior of 
solution (\ref{k=2solution1}) 
in the region 
$\Delta^{(1)}\ll \eta^{1/2}$. 
It turns out to be 
\begin{eqnarray}
N&=& 
\left( \begin{array}{cc} 
  z_1^{(0)} & 
     \sqrt{\eta} z_1^{(1)}/\Delta^{(1)} 
\\ 
  0 & z_1^{(0)} 
\end{array} \right)
+
 \left( \begin{array}{cc} 
   z_1^{(1)}/2 & 0 
\\ 
  0 & -z_1^{(1)}/2 
\end{array} \right)
+O(\Delta^{(1)2}/\eta),    
\nonumber \\
M&=& 
\left( \begin{array}{cc} 
  z_2^{(0)} & 
     \sqrt{\eta}z_2^{(1)}/\Delta^{(1)} 
\\ 
  0 & z_2^{(0)} 
\end{array} \right)
+
 \left( \begin{array}{cc} 
    z_2^{(1)}/2 & 0 
\\ 
  0 & -z_2^{(1)}/2 
\end{array} \right)
+O(\Delta^{(1)2}/\eta),    
\nonumber \\
H_{\dot{1}}&=& 
\left( \begin{array}{c} 
0 \\
 \sqrt{2\eta} \end{array}\right)
+
\left(\begin{array}{c} 
    \Delta^{(1)}/\sqrt{2} \\
  0 
\end{array} \right) 
+O(\Delta^{(1)2}/\eta). 
\label{k=2solution2}
\end{eqnarray} 
It tells that, 
under the limit of two D5-branes  
colliding  with each other, 
solution (\ref{k=2solution1}) 
does depend on how they approach. 
It really depends on the way that 
$z^{(1)}=(z^{(1)}_1,z^{(1)}_{2})$ 
goes to $(0,0)$. 
It is parametrized by 
$\lambda^{(1)}
=(\lambda^{(1)}_1,\lambda^{(1)}_2)$ : 
\begin{eqnarray}
(\lambda^{(1)}_1,\lambda^{(1)}_2)
=\lim_{z^{(1)}\rightarrow (0,0)} 
(z^{(1)}_1/\Delta^{(1)},z^{(1)}_2/\Delta^{(1)}). 
\label{lambda1}
\end{eqnarray}
Notice that $\lambda^{(1)}\in S^3$. 
With this parametrization of their collision 
the solution of ADHM 
equation acquires the following form 
when two D5-branes overlap  : 
\begin{eqnarray}
N^{(0)}
&=& \left( \begin{array}{cc}
  z^{(0)}_{1} &  
       \sqrt{\eta}\lambda^{(1)}_{1} \\
   0  &  z^{(0)}_{1} 
         \end{array} \right) ,       
\nonumber   \\ 
M^{(0)}
&=& 
\left( \begin{array}{cc}
  z^{(0)}_{2} &  
       \sqrt{\eta}\lambda^{(1)}_{2} \\
   0  &  z^{(0)}_{2} 
         \end{array} \right),  
\nonumber \\ 
H^{(0)}_{\dot{1}}
&=& 
\left( \begin{array}{c} 
  0 \\ \sqrt{2\eta} 
         \end{array} \right),~~~~ 
H^{(0)}_{\dot{2}}
= \left( \begin{array}{c} 
  0 \\ 0 
         \end{array} \right).     
\label{k=2solution3}
\end{eqnarray}

            Although solution (\ref{k=2solution3})
has the collision parameter $\lambda^{(1)}$ of 
two D5-branes besides the position $z^{(0)}$ 
where they overlap, different collisions of 
two D5-branes at $z^{(0)}$ do not necessarily 
correspond to different points in  the moduli space 
${\cal M}^{k=2}$ due to the symmetry enhancement. 
In fact, let us consider 
the action of the $U(1)$ subgroup,   
$\left\{ 
  \left( \begin{array}{cc}
          e^{i \theta} & 0 \\
                  0 & 1 
                     \end{array} \right)  
  \in U(2) \right\}$. 
Solution (\ref{k=2solution3}) 
transforms to : 
$\left( g=\left( \begin{array}{cc}
          e^{i \theta} & 0 \\
                  0 & 1 
                     \end{array} \right)  \right)$
\begin{eqnarray} 
N^{(0) ~g}
&=& \left( \begin{array}{cc}
  z^{(0)}_{1} &  
       \sqrt{\eta}\lambda^{(1)}_{1}e^{i \theta} \\
   0  &  z^{(0)}_{1} 
         \end{array} \right) ,       
\nonumber   \\ 
M^{(0)~g}
&=& 
\left( \begin{array}{cc}
  z^{(0)}_{2} &  
       \sqrt{\eta}\lambda^{(1)}_{2}e^{i \theta} \\
   0  &  z^{(0)}_{2} 
         \end{array} \right),  
\label{k=2U(1)trans}
\end{eqnarray}
with $H^{(0)}_{\dot{A}}$ unchanged,~ 
$H^{(0)~g}_{\dot{A}}
=H^{(0)}_{\dot{A}}$. 
Only the collision parameter 
$\lambda^{(1)}=(\lambda^{(1)}_1,\lambda^{(1)}_2)$ 
is multiplied by $e^{i \theta}$. 
This multiplication by the $U(1)$ phase factor 
means that  
we can distinguish overlapping two D5-branes by 
its collision parameter $\lambda^{(1)}$ only up to 
the $U(1)$ action. Taking it in reverse, 
in the projection map (\ref{pi}), 
the inverse image of the configuration 
of two D5-branes degenerate at $P \in {\bf R}^4$ 
is equivalent 
to $S^2$ which is obtainable by modding out 
$S^3$ by $U(1)$ : 
$\pi^{-1}(2P) \cong S^2$. 
By looking at solution (\ref{k=2solution3}) 
we find out that the area of this $S^2$ 
is proportional to $\eta (>0)$ and, 
as we shall see later,  
this cycle can be regarded as 
one of cousins of the bound states of D5-branes.

            At this stage it may be convenient to reinterpret 
these phenomena from  
the world-volume $U(2)$ gauge theoretical point of view. 
Notice that $U(2)$ gauge symmetry is completely broken 
by Higgs mechanism 
when the positions of two D5-branes are generic 
($z^{(1)}\neq (0,0)$). 
At the point where two D5-branes overlap each other 
only the $U(1)$ gauge symmetry is restored 
and other gauge symmetries are still 
left broken due to $\eta (>0)$.  
By taking the parametrization 
\begin{eqnarray}
(z^{(1)}_1,z^{(1)}_2)=
(\Delta^{(1)}cos \beta e^{i\sigma^{(1)}_1},
\Delta^{(1)}sin \beta e^{i\sigma^{(1)}_2}), 
\label{k=2radial}
\end{eqnarray}
we can find that 
$\sigma^{(1)}_{U(1)}=\frac{1}{2}(\sigma^{(1)}_1+
\sigma^{(1)}_2)$ is the corresponding $U(1)$ 
Nambu-Goldstone boson.

\subsubsect{$(ii)~~k \geq 3~~case$ }

                 Let us consider the situation that 
$k$ D5-branes overlap at $P \in {\bf R}^4$. 
It is very similar to the case of $k=2$. 
The $U(k)$ gauge symmetry on the worldvolume theory 
is completely broken by Higgs mechanism 
when these $k$ D5-branes 
are in generic positions. 
At the point where 
they degenerate occurs the 
restoration of 
$\underbrace{U(1)\times \cdots \times U(1)}_{k-1}$ 
gauge symmetry and other gauge symmetries are still 
left broken due to $\eta (>0)$.

          To see explicitly,  
let $k$ D5-branes collide at 
$P$ with keeping their generic positions. As in the case 
of $k=2$ 
we shall introduce the collision parameters 
$\lambda^{(j)}=(\lambda^{(j)}_1,\lambda^{(j)}_2)
\in S^3$ from the phases of relative coordinates 
$z^{(j)}=(z^{(j)}_1,z^{(j)}_2)$ ($1 \leq j \leq k-1$). 
These collision parameters will appear in the solution 
of ADHM equation which is obtainable as the limit of 
the $k$-ply degeneration of D5-branes. 
Even though they survive in the solution 
we can not necessarily distinguish 
all the different collisions at $P$ 
due to the enhanced symmetry group 
$\underbrace{U(1)\times \cdots \times U(1)}_{k-1} $. 
In particular the net degrees of freedom of $\lambda^{(j)}$s 
are $3(k-1)-(k-1)=2(k-1)$. 
It means 
that  
in projection map (\ref{pi}), 
$\pi^{-1}(kP)$,  the inverse image of 
the configuration of 
$k$ D5-branes overlapping at $P$,  
is the cycle of $2(k-1)$ dimensions.

       To give an exact description of this cycle, 
it is necessary to study the $k$-ply 
degenerations of $k$ D5-branes 
over all the patterns of their collisions. 
These patterns 
may be parametrized by Young tableaux 
with length $k$. 
Namely, by letting $Y=\mbox{[}a_1,\cdots,a_l \mbox{]}$ 
be a Young tableau with length $k$ 
$~(a_1 \geq \cdots \geq a_l \geq 1,~
|Y| \equiv a_1+\cdots +a_l
=k~)$,  
the corresponding pattern of collisions is that 
$a_j$ 
$(1 \leq j \leq l)$ of $k$ D5-branes 
degenerate first 
and then they collide at $P$ in order. 
From this observation one may 
say that the stratification of 
the cycle $\pi^{-1}(kP)$ can be 
labelled by these Young tableaux :  
$\pi^{-1}(kP)=
   \coprod_{|Y|=k}\pi^{-1}(kP)_{Y}$.  
The situation we have mentioned above includes the maximal 
stratum $\pi^{-1}(kP)_{Y=\mbox{[}k \mbox{]}}$.

\subsect{Construction of Observables}

       Let us construct physical observables 
in our theory. 
Due to BRST symmetry 
(\ref{BRST}) 
the physical content of the theory will 
become equivalent to the cohomology theory of the moduli 
space ${\cal M}^k$. Especially the BRST charge 
$Q_{BRST}$ can be regarded as the differential operator 
$d_{{\cal M}^k}$ on the moduli space :
\begin{eqnarray}
Q_{BRST}=d_{{\cal M}^k} ~~:~~
\Omega^*({\cal M}^k) \rightarrow 
\Omega^*({\cal M}^k) .
\end{eqnarray}

\subsubsect{$(i)~~ k=1~~case$}

         Notice that ${\cal M}^{k=1}=$ ${\bf R}^4$. 
$X_{A \dot{A}}$ gives the global coordinates of the moduli 
space. 
There is an unique physical observable $\mbox{[}e\mbox{]}$ 
$\in H^0({\bf R}^4)$. 
We shall denote it by $O_1$ : 
\begin{eqnarray}
O_1=\mbox{[}e\mbox{]}. 
\label{O1}
\end{eqnarray}

\subsubsect{$(ii)~~k \geq 2~~case$}

              For the case of $k \geq 2$ 
our topological theory  
will be effectively described  
as the sum of two topological subsystems. 
One is the system of $x^{(0)}_{A\dot{A}}$, 
which leads to the topological quantum mechanics on 
${\bf R}^4$. The BRST charge $Q_{BRST}$ acts 
as the differential operator $d_{{\bf R}^4}$ on ${\bf R}^4$.
So there is an unique physical observable 
$\mbox{[}e\mbox{]} \in H^0({\bf R}^4)$ in this subsystem. 
The another is the system of relative coordinates 
$x^{(j)}_{A\dot{A}}$ $(1 \leq j \leq k-1)$, which gives us 
the nontrivial physical observables.

       Let us consider the following combinations of 
field variables : 
\begin{eqnarray}
O_k 
\equiv 
\mbox{[ $e$ ]} 
\prod_{j=1}^{k-1}
\frac{2}{\sqrt{\pi \rho}} 
e^{-\frac{\Delta^{(j)~2}}{\rho}}
\psi^{(j)}_{\Delta} 
\psi^{(j)}_{U(1)}, 
\label{Ok}
\end{eqnarray}
where $\rho$ is a positive constant 
and $\mbox{[ $e$ ]}$ is the physical 
observable of the subsystem of 
$x^{(0)}_{A\dot{A}}$. 
$\psi^{(j)}_{\Delta}$  
is the BRST partner of the $j$-th relative distance 
$\Delta^{(j)}$ 
\begin{eqnarray} 
\delta_{\bf b}\Delta^{(j)}=
\psi^{(j)}_{\Delta}  
~~~~~(1 \leq j \leq k-1), 
\end{eqnarray}  
and $\psi^{(j)}_{U(1)}$ is 
the BRST partner of 
the $j$-th Nambu-Goldstone boson 
$\sigma^{(j)}_{U(1)}$  
\begin{eqnarray}
\delta_{\bf b}\sigma^{(j)}_{U(1)}
=\psi^{(j)}_{U(1)}~~~~~
(1 \leq j \leq k-1).
\end{eqnarray}  
$O_k$ is the physical observable with 
ghost number equal to $(1+1)\times (k-1)=2(k-1)$.  
Let us investigate its geometrical meaning. 
Due to the exponential factors in (\ref{Ok}) 
the support of $O_k$ 
in the moduli space ${\cal M}^k$ 
is on a tubular 
neighborhood (with it's radius $\sim \sqrt{\rho}$) 
of the cycle 
${\cal M}^{k}_{(k)}
\equiv 
\{ \pi^{-1}(kP) : P \in {\bf R}^4 \}$. 
The dimensions of this cycle are 
$2(k-1)+4=2(k+1)$ 
(or the co-dimensions equal $2(k-1)$). 
Ghost fields $\psi^{(j)}_{\Delta}$ and 
$\psi^{(j)}_{U(1)}$ appearing in (\ref{Ok}) 
can be regarded as the differential forms 
normal to this $2(k+1)$-dimensional cycle. 
Therefore $O_k$ can be interpreted as the 
Poincar\'{e} dual of the cycle ${\cal M}^k_{(k)}$.

    Besides the above 
understanding of $O_k$ based on  
the geometry of moduli space ${\cal M}^k$
it is quite 
surprising that $O_k$ itself has its own 
meaning in any topological system 
based on  the worldvolume gauge theory 
with gauge group 
$U(k')$ ($k' \geq k$). 
In fact, 
suppose that there exist $k'$ D5-branes and 
$k$ pieces of them are colliding at a point. 
As regards these degenerate $k$ D5-branes 
we can apply the preceding discussions and then  
we will obtain $O_k$ as the physical observable. 
Due to this universality 
\footnote{This kind of universality is familiar  
in the non-perturbative formulation 
of two-dimensional quantum gravity. 
Namely the gravitational descendents can be defined 
non-perturbatively and also have 
their geometrical expressions  
on the moduli space of Riemann surface with 
fixed genus.} 
$O_k$ will be 
regarded as 
a field theoretical realization of 
the \lq\lq correspondences"  given 
in \cite{NAKAJIMA},\cite{GROJNOWSKI}.

         Although one may ask the possibility 
that there exist physical observables other than 
$O_k$, we will see in the next subsection that 
these $O_k$ generate all the observables.

\subsect{Second-Quantized D5-Branes}

        In order to study the implication of 
this universality  
of physical observables 
let us consider the quantization of the theory 
by means  
of operator formalism.

         Since the physical observables 
$O_k$ $(k \geq 1)$ have 
the universality, the vacuum state should be 
characterized 
in terms of D5-branes, not by gauge theories.  
It may be very plausible to introduce the vacuum as the 
state which simply corresponds to the configuration of 
{\em no} D5-branes. Let us denote it by 
$|0 \rangle$.   
We shall introduce the physical operator 
$\alpha_{-k}$ 
which corresponds to $O_k$. 
From the look of the explicit form of $O_k$ 
it is the bosonic operator with 
ghost number equal to $2(k-1)$ and will be 
interpreted as the creation operator 
of the configuration of 
$k$ D5-branes degenerating at a  
point. One may also say that 
$\alpha_{-k}$ is the creation  
operator of the bound state of $k$ D5-branes.
Notice that, 
because $O_k$ can be regarded as   
the Poincar\'e dual of an appropriate  
cycle in each moduli space ${\cal M}^{k'}$ 
($k' \geq k$), 
this identification of $\alpha_{-k}$ 
with the creation operartor of the bound state  
also gives the geometrical interpretation of 
the bound states of D5-branes. 
Stability of these bound 
states will be ensured topologically by their 
identification with the cycles of the moduli spaces. 
Since the creations of D5-branes should not 
depend on their order these  operators will 
commute one another : 
$\mbox{[~$\alpha_{-k_1}$,~$\alpha_{-k_2}$~]}=0$. 
Being given the creation operators it is  
also conceivable to consider the annihilation of 
these bound states  
\footnote{The correspondences are given in 
\cite{NAKAJIMA},\cite{GROJNOWSKI}.}. 
The annihilation operator 
$\alpha_k (=\alpha_{-k}^{\dagger})$ 
of the bound state of 
$k$ D5-branes will be the bosonic 
operator of ghost number equal to $-2(k-1)$ 
satisfying the commutation relations,  
\begin{eqnarray}
 \mbox{[}~\alpha_{m},~\alpha_{n} \mbox{]}
=m \delta_{m+n,~0}~~~~~~
(m,n \in {\bf Z}\backslash \{0\}) .
\label{CR1}
\end{eqnarray}

                Being restricted to the $U(k)$ 
gauge theory the basis of the physical 
Hilbert space 
${\cal H}^k$ will have the form,  
\begin{eqnarray}
\alpha_{-k_1}\cdots \alpha_{-k_l}
|0 \rangle~~~~~~~~ 
(\forall k_i \geq 1,~k_1+\cdots+k_l=k),  
\end{eqnarray}
which is the bosonic state with ghost number 
equal to $2(k-l)$. 
Because of this restriction on the allowed 
combinations of the creation operators 
it becomes quite reasonable to 
introduce the total Hilbert space ${\cal H}$ :  
\begin{eqnarray}
{\cal H} \equiv 
\oplus_{k=0}^{\infty} {\cal H}^k, 
\label{totalH}
\end{eqnarray}
and it is very tempting to interpret 
this total Hilbert space ${\cal H}$ 
as the physical Hilbert space of 
the second-quantized D5-branes 
(or string-solitons ). 
Let us investigate 
this possibility furthur. 
Since, by taking mass of a D5-brane as an unit, 
the bound state energy of $k$ D5-branes is 
$k$, we shall introduce 
the Hamiltonian operator $\hat{H}$ 
of the second-quantized D5-branes as follows : 
\begin{eqnarray}
\hat{H}=\sum_{k \geq 1}\alpha_{-k}\alpha_{k}. 
\label{energy-op}
\end{eqnarray} 
We also denote the ghost number operator by 
$\hat{N}$. Note that $\hat{N}$ measures 
degrees of the Poincar\'e duals of the cycles 
which geometrically realize the 
bound states of D5-branes. 
One may regard 
$\mbox{Tr}_{\cal{H}}t^{\hat{N}}q^{\hat{H}}$ 
as an analogue of the partition function 
of these  second-quantized D5-branes 
(or string solitons). 
It has the form 
\begin{eqnarray}
\mbox{Tr}_{\cal{H}}t^{\hat{N}}q^{\hat{H}}
=\frac{1}
  {\prod_{l\geq 1}(1-t^{2(l-1)}q^l) }, 
\end{eqnarray}
which was shown to be equal to 
$\sum_{k\geq 0}q^kP_t({\cal M}^k)$ 
\cite{NAKAJIMA}. 
($P_t({\cal M}^k)$ is the Poincar\'e 
polynomial of ${\cal M}^k$.) 
This coincidence of the partition function with 
the generating function of the Poincar\'e 
polynomials also shows that 
$\{ O_k \}_{k \geq 1}$ generate all the physical observables  
of our topological field theory.

\subsect{Generalization}

          In the preceding discussions our study 
was limitted to the case when D5-branes are on ${\bf R}^4$. 
Let us generalize it to the case that they are on a curved 
background $X$. 
The configuration space of $k$ D5-branes is the $k$-th 
symmetric product of $X$, $S^k(X)$. 
We may need to introduce the moduli space 
${\cal M}^k(X)$, that is, 
an analogue of 
${\cal M}^k (\equiv {\cal M}^k({\bf R}^4))$ . 
The moduli space ${\cal M}^k(X)$ will give a 
resolution of $S^k(X)$ as 
${\cal M}^k({\bf R}^4)$ does give 
a resolution of $S^k({\bf R}^4)$. 
In particular those cycles which appear by the 
resolution will lead to the geometrical realization 
of the bound states of D5-branes on $X$. 
Since the singularities of $S^k(X)$ are those configurations 
that some of D5-branes are overlapping at points of $X$, 
their resolution will be prescribed by the local 
properties of $X$ in the neighborhood of these points, and 
therefore will be quite similar to the case of ${\bf R}^4$ 
or ${\bf C}^2$. 
For the case that $X$ is an algebraic surface the moduli 
space ${\cal M}^k(X)$ is known \cite{NAKAJIMA} 
to be given by $X^{ \left[ k \right]}$, 
the Hilbert scheme  of $k$ points of $X$ .

                Due to this locality our framework of 
the second quantization of D5-branes will practically work  
even in this case. 
Let us consider the same theory as before. 
The effect of the curved background will be taken into 
account by modifying the subsystem of the center of 
D5-branes to the topological quantum mechanics on $X$. 
Although the division of the system into two parts, 
that is, two subsystems of the center and the relative motions  
of D5-branes seems to be meaningless in the curved space $X$, 
it will work at least 
for the construction of physical observables. 
This is because, as we have seen, 
$O_k$ is essentially zero except for the configurations 
that D5-branes are overlapping at a point.

                Since the topological quantum mechanics on $X$ 
is equivalent to the cohomology theory of $X$, 
by applying the same construction as that of 
$O_k(\mbox{[$e$]}) (\equiv O_k)$, 
each element $\mbox{[$\omega$]}$ of 
$H^*(X)$ gives us the physical observable 
$O_k(\mbox{[$\omega$]})$. 
It is bosonic (fermionic) when 
$d_{\omega}$, 
the degrees of $\omega$, is even (odd). 
It has the ghost number $2(k-1)+d_{\omega}$. 
Let us also introduce the creation operator 
$\alpha_{-k}(\mbox{[$\omega$]})$ corresponding to 
the physical observable  $O_k(\mbox{[$\omega$]})$ 
and then we will identify it 
with the creation operator of 
the bound state of $k$ D5-branes topologically constrained on 
the cycle $C$ of $X$. 
($\mbox{[$\omega$]}$ is the Poincar\'e dual of $C$ in $X$.)  
These creation operators of the bound states 
constrained on the even (odd) dimensional cycles  of $X$  
constitute (anti-)commuting bosonic (fermionic) operators. 
The total physical Hilbert space 
${\cal H}$ can be defined as the Fock 
space of these creation operators.  
Therefore $\mbox{Tr}_{{\cal H}}t^{\hat{N}}q^{\hat{H}}$ 
has the form 
\begin{eqnarray}
\mbox{Tr}_{{\cal H}}t^{\hat{N}}q^{\hat{H}}
=\frac{\prod_{l\geq 1}(1+t^{2l-1}q^l)^{b_1(X)}
              (1+t^{2l+1}q^l)^{b_3(X)}}
{\prod_{l\geq 1}
   (1-t^{2(l-1)}q^l)^{b_0(X)}
         (1-t^{2l}q^l)^{b_2(X)}
             (1-t^{2(l+1)}q^l)^{b_4(X)}}, 
\label{partition on X}
\end{eqnarray} 
where $b_i(X)$ is the $i$-th Betti number of $X$. 
For the case that $X$ is a projective surface, 
it coincides \cite{NAKAJIMA} 
with $\sum_{k \geq 0}q^k P_t(X^{\left[ k \right]})$, 
the generating function of the Poincar\'e 
polynomials of the Hilbert scheme parametrizing 
points in $X$ 
\cite{GOTTSCHE}. 
In particular, by setting $t=-1$, 
$\mbox{Tr}_{{\cal H}}(-)^{\hat{N}}q^{\hat{H}}$ 
(\ref{partition on X}) reduces to the generating 
function of the Euler numbers and coincides with 
the partition function of 
$N=4$ supersymmetric Yang-Mills theory proposed by 
Vafa-Witten \cite{VAFA-WITTEN}.

\sect{Discussion}

\subsubsect{{\em CFT of second quantized D5-branes}}

              The second quantization of 
D5-branes described in this article may be considered 
as the quantization in terms of two-dimensional conformal 
field theory 
\footnote{The creation and annihilation operators 
$\alpha_n$ of the bound states of D5-branes, 
with an addition of the zero-mode, 
constitute a free boson 
$\partial \phi(z) \equiv \sum_n \alpha_{n}z^{-n-1}$. }. 
This unexpected appearance of $2D$ CFT may be the origin 
of the integrability of four-dimensional $N=4,2$ supersymmetric 
Yang-Mills theories. 
The Seiberg-Witten solutions 
\cite{SEIBERG-WITTEN}, that is, 
the exact low energy effective actions of these 
supersymmetric theories are prescribed by introducing the moduli 
of the curves characteristic of these theories.   
It is also pointed out 
\cite{Integrability}  
that these exact solutions can be regarded as the semi-classical 
solutions of two dimensional integrable systems such as 
the Toda lattice \cite{UENO-TAKASAKI}.  
Notice that 
these integrable system can be also realized  
\cite{JIMBO-MIWA-SATO} by using $2D$ CFT. 
Since these exact solutions of $4D$ supersymmetric 
Yang-Mills theories should be 
ultimately obtainable from 
string theory 
these two appearances of $2D$ CFT will not be 
accidental.

           An attempt to understand the Seiberg-Witten exact 
solutions from the $2D$ CFT viewpoint has been made in 
\cite{TAKASAKI-NAKATSU}, where, 
by emphasizing the similarity with the non-perturbative 
formulation of two-dimensional gravity
\footnote{The string equation, $\mbox{[$P,Q$]}=1$, 
of $2D$ gravity defines an isomonodromy 
problem \cite{MOORE}. }, 
isomonodromic deformation problems 
are addressed so that 
their semi-classical analysis precisely gives us the 
Seiberg-Witten solutions. 
One may say that our framework of the second 
quantization of D5-branes is on the line 
suggested by the authors of \cite{TAKASAKI-NAKATSU} 
because the quantization follows from the universality 
of physical observables which, as we mentioned 
in the text, is also characteristic in 
the nonperturbative formulation of two-dimensional 
gravity.  
In order to step furthur it might be necessary to 
give a more precise treatment of the second quantization of 
D5-branes (presumably without using any gauge theories). 
The construction of matrix models which 
realize the correspondences given in 
\cite{NAKAJIMA},\cite{GROJNOWSKI} should be investigated.

\newpage 

\noindent
{\Large\bf Acknowledgements} 
\vspace{5mm}

We are very grateful to  
H.Nakajima for helpful explanations of his works 
related on this topics.  
We also benefited from discussions with 
K. Takasaki, Y.Yasui and H.Kanno. 
Last, but not least, 
we would like to thank Y.Hashimoto for 
numerous discussions and explanations on  
several mathematical points. 
This work is supported in part by 
Grants-in-Aid for 
Scientific Research (08640372, 08304001) from 
the Ministry of Education, Science and Culture, 
Japan.


\newpage 

\vspace{20mm}
\begin{center}

\begin{tabular}{|c|c|c|c|}\hline
fields & matrix size & statistics & ghost number  \\ \hline
\hline
$A_{\mu}$ & $k\times k$ & bose & 0 \\
$X_i$ & $k\times k$ & bose & 0 \\
$H_{\dot{A}}$ & $k\times n$ & bose & 0 \\
$\psi_{\mu}$ & $k\times k$ & fermi & 1 \\
$\psi_i$ & $k\times k$ & fermi & 1 \\
$\psi_{\dot{A}}$ & $k\times n$ & fermi & 1 \\
$\eta_{\mu\nu}$ & $k\times k$ & fermi & -1 \\
$\eta_{\mu i}$ & $k\times k$ & fermi & -1 \\
$\eta$ & $k\times k$ & fermi & -1 \\
$\eta_{\mu\dot{A}}$ & $k\times n$ & fermi & -1 \\
$\eta_{\dot{A}\dot{B}}$ & $k\times k$ & fermi & -1 \\
$\chi_{\mu\nu}$ & $k\times k$ & bose & 0 \\
$\chi_{\mu i}$ & $k\times k$ & bose & 0 \\
$\chi_{\mu\dot{A}}$ & $k\times n$ & bose & 0 \\
$D_{\dot{A}\dot{B}}$ & $k\times k$ & bose & 0 \\
$\phi$ & $k\times k$ & bose & 2 \\
$\lambda$ & $k\times k$ & bose & -2 \\ \hline
\end{tabular}

\vspace{5mm}
Table 1:
The field contents of the topological field theory.   
\lq\lq $k \times k$" (\lq\lq $k \times n$") 
in the above means that these fields are 
$k \times k$ hermitian ($k \times n$ complex) 
matrix-valued.
\vspace{10mm}
\end{center}


\end{document}